\documentclass[conference]{IEEEtran}
\IEEEoverridecommandlockouts

\usepackage{amsmath}
\usepackage{amsfonts}
\usepackage{amssymb}
\usepackage{graphicx}
\usepackage{epsfig}
\usepackage{subfig}
\usepackage{float}
\usepackage{epstopdf}
\usepackage{array}
\usepackage{multirow}
\usepackage{setspace}
\usepackage{color}
\usepackage[section]{placeins}
\usepackage{url}
\usepackage{algorithm}
\usepackage{algpseudocode}
\usepackage{multicol}
\usepackage{lipsum}
\usepackage{soul}
\usepackage{comment}
\usepackage{balance}
\usepackage{xcolor}
\usepackage{cite}
\usepackage[normalem]{ulem}
\usepackage[acronym]{glossaries}

\title{Reducing the Paging Overhead in Highly Directional Systems}

\author{
  \IEEEauthorblockN{
   Sanjay Goyal, Hussain Elkotby, Ravikumar Pragada, and Tanbir Haque
   }
    
	\IEEEauthorblockA{InterDigital Communications, Inc., USA \\
	emails:
	\{sanjay.goyal, hussain.elkotby, ravikumar.pragada, tanbir.haque\}@interdigital.com 
	}
}

\begin{document}
\maketitle

\begin{abstract}
New Radio (NR) supports operations at high-frequency bands (e.g., millimeter-wave frequencies) by using narrow beam based directional transmissions to compensate high propagation losses at such frequencies. Due to the limited spatial coverage with each beam, the broadcast transmission of paging in NR is performed using beam sweeping, which takes multiple time slots. Thus, the paging procedure used in NR would substantially increase the downlink resource overhead of the network with directional transmissions. Such overhead would further increase as we move higher in the frequency bands, such as terahertz bands, which is being viewed as one of the potential candidates for future generation networks. Therefore, the NR based paging solution is infeasible for supporting highly directional systems. In this paper, we propose a novel minimal feedback enabled paging mechanism, which instead of using all the beams for paging transmissions, only activates sub-set of beams having one or more UEs under the coverage. UE presence indications are implemented to identify the correct set of beams to be activated. Our analytical analysis and simulations show that the proposed solution significantly reduces the downlink paging overhead compared to the NR based solution (e.g., more than 80\% gain for a system supporting 64 number of beams at a UE density of 200 UEs per paging occasion) while incurring minimal energy cost at the UE side.

\end{abstract}

\IEEEpeerreviewmaketitle

\newacronym{tx}{TX}{Transmission}
\newacronym{po}{PO}{Paging Occasion}
\newacronym{scs}{SCS}{SubCarrier Spacing}
\newacronym{ofdm}{OFDM}{Orthogonal Frequency Division Multiplexing}
\newacronym{fdm}{FDM}{Frequency Division Multiplexing}
\newacronym{cpu}{CPU}{Central Processing Unit}
\newacronym{tcp}{TCP}{Transmission Control Protocol}
\newacronym{tcpw}{TCPW}{TCP Wave}
\newacronym{lte}{LTE}{Long Term Evolution}
\newacronym{nr}{NR}{New Radio}
\newacronym{nru}{NR-U}{NR-based access to Unlicensed spectrum}
\newacronym{cca}{CCA}{Clear Channel Assessment}
\newacronym{cwnd}{cWnd}{Congestion Window}
\newacronym{caa}{CAA}{Congestion Avoidance Algorithm}
\newacronym{bbr}{BBR}{Bottleneck Bandwidth and Round-trip propagation time}
\newacronym{nv}{NV}{New Vegas}
\newacronym{rtt}{RTT}{Round-Trip Time}
\newacronym{ietf}{IETF}{Internet Engineering Task Force}
\newacronym{rfc}{RFC}{Request For Comments}
\newacronym{gcc}{GCC}{GNU Compiler Collection}
\newacronym{tso}{TSO}{TCP Segmentation Offloading}
\newacronym{tsq}{TSQ}{TCP Small Queue}
\newacronym{gbr}{GBR}{Guaranteed Bit Rate}
\newacronym{nongbr}{non-GBR}{non-Guaranteed Bit Rate}
\newacronym{enb}{eNB}{Evolved Node B}
\newacronym{dpi}{DPI}{Deep Packet Inspection}
\newacronym{rlc}{RLC}{Radio Link Control}
\newacronym{bsr}{BSR}{Buffer Status Report}
\newacronym{qos}{QoS}{Quality of Service}
\newacronym{aqm}{AQM}{Active Queue Management}
\newacronym{rds}{RDS}{Radio Data Scheduler}
\newacronym{tc}{TC}{Traffic Control}
\newacronym{drb}{DRB}{Data Radio Bearer}
\newacronym{rnti}{RNTI}{Radio Network Temporary Identifier}
\newacronym{bql}{BQL}{Byte Queue Limits}
\newacronym{ue}{UE}{User Equipment}
\newacronym{am}{AM}{Acknowledged Mode}
\newacronym{epc}{EPC}{Evolved Packet Core}
\newacronym{cn}{CN}{Core Network}
\newacronym{gnb}{gNB}{next-Generation Node B}
\newacronym{ran}{RAN}{Radio Access Network}
\newacronym{3gpp}{3GPP}{3rd Generation Partnership Project}
\newacronym{5g}{5G}{fifth Generation}
\newacronym{dl}{DL}{DownLink}
\newacronym{ul}{UL}{UpLink}
\newacronym{tti}{TTI}{Transmission Time Interval}
\newacronym{sr}{SR}{Scheduling Request}
\newacronym{bsr}{BSR}{Buffer Status Report}
\newacronym{e2e}{E2E}{End-To-End}
\newacronym{embb}{eMBB}{enhanced Mobile BroadBand}
\newacronym{urllc}{URLLC}{Ultra-Reliable and Low-Latency Communications}
\newacronym{mmtc}{mMTC}{massive Machine Type Communications}
\newacronym{ul}{UL}{UpLink}
\newacronym{dl}{DL}{DownLink}
\newacronym{phy}{PHY}{Physical layer}
\newacronym{mac}{MAC}{Medium Access Control}
\newacronym{prb}{PRB}{Physical Resource Block}
\newacronym{pps}{pps}{Packets Per Second}
\newacronym{cp}{CP}{Cyclic Prefix}
\newacronym{tbs}{TBS}{Transport Block Size}
\newacronym{tb}{TB}{Transport Block}
\newacronym{cb}{CB}{Code Block}
\newacronym{gtp}{GTP}{GPRS Tunneling Protocol}
\newacronym{sap}{SAP}{Service Access Point}
\newacronym{tm}{TM}{Transparent Mode}
\newacronym{um}{UM}{Unacknowledged Mode}
\newacronym{am}{AM}{Acknowledged Mode}
\newacronym{sm}{SM}{Saturation Mode}
\newacronym{sinr}{SINR}{Signal-to-Interference-plus-Noise Ratio}
\newacronym{rrc}{RRC}{Radio Resource Control}
\newacronym{tdma}{TDMA}{Time-Division Multiple Access}
\newacronym{ofdma}{OFDMA}{Orthogonal Frequency-Division Multiple Access}
\newacronym{rbg}{RBG}{Resource Block Group}
\newacronym{rb}{RB}{Resource Block}
\newacronym{dci}{DCI}{Downlink Control Information}
\newacronym{uci}{UCI}{Uplink Control Information}
\newacronym{ipat}{IPAT}{Inter-Packet Arrival Time}
\newacronym{pdsch}{PDSCH}{Physical Downlink Shared Channel}
\newacronym{pusch}{PUSCH}{Physical Uplink Shared Channel}
\newacronym{pucch}{PUCCH}{Physical Uplink Control Channel}
\newacronym{pdcch}{PDCCH}{Physical Downlink Control Channel}
\newacronym{tdd}{TDD}{Time Division Duplex}
\newacronym{fdd}{FDD}{Frequency Division Duplex}
\newacronym{rach}{RACH}{Random Access Channel}
\newacronym{cbr}{CBR}{Constant Bit Rate}
\newacronym{los}{LoS}{Line-of-Sight}
\newacronym{mcs}{MCS}{Modulation Coding Scheme}
\newacronym{bwp}{BWP}{Bandwidth Part}
\newacronym{cqi}{CQI}{Channel Quality Indicator}
\newacronym{bler}{BLER}{Block Error Rate}
\newacronym{tbler}{TBLER}{Transport Block Error Rate}
\newacronym{mi}{MI}{Mutual Information}
\newacronym{l2sm}{L2SM}{Link to System Mapping}
\newacronym{sliv}{SLIV}{Start and Length Indicator Value}
\newacronym{mmwave}{mmWave}{millimeter-wave}
\newacronym{pdu}{PDU}{Packet Data Unit}
\newacronym{ca}{CA}{Carrier Aggregation}
\newacronym{snr}{SNR}{Signal-to-Noise Ratio}
\newacronym{sinr}{SINR}{Signal to Interference-plus-Noise Ratio}
\newacronym{pdcp}{PDCP}{Packet Data Convergence Protocol}
\newacronym{sdap}{SDAP}{Service Data Adaptation Protocol}
\newacronym{sdu}{SDU}{Service Data Unit}
\newacronym{nas}{NAS}{Non-Access Stratum}
\newacronym{sme}{SME}{Small and Medium Enterprise}
\newacronym{rat}{RAT}{Radio Access Technology}
\newacronym{pgw}{PGW}{Packet data network GateWay}
\newacronym{sgw}{SGW}{Service GateWay}
\newacronym{ldpc}{LDPC}{low Density Parity Check}
\newacronym{cca}{CCA}{Clear Channel Assessment}
\newacronym{csmaca}{CSMA/CA}{Carrier Sense Multiple Access with Collision Avoidance}
\newacronym{ccm}{CCM}{Component Carrier Manager}
\newacronym{cam}{CAM}{Channel Access Manager}
\newacronym{cws}{CWS}{Contention Window Size}
\newacronym{ed}{ED}{Energy detection}
\newacronym{harq}{HARQ}{Hybrid Automatic Repeat Request}
\newacronym{ap}{AP}{Access Point}
\newacronym{laa}{LAA}{Licensed-Assisted Access}
\newacronym{lbt}{LBT}{Listen-Before-Talk}
\newacronym{mcot}{MCOT}{Maximum Channel Occupancy Time}
\newacronym{cot}{COT}{Channel Occupancy Time}
\newacronym{rar}{RAR}{Random Access Response}
\newacronym{ip}{IP}{Internet Protocol}
\newacronym{ss/pbch}{SS/PBCH}{Synchronization Signal/Physical Broadcast Channel}
\newacronym{prach}{PRACH}{Physical Random Access Channel}

\section{Introduction} 
\label{sec:intro}
\gls{nr}, the 3rd Generation Partnership Project (3GPP) Radio Access Network (RAN) for the fifth-generation (5G) cellular networks, has a flexible, scalable, and forward-compatible design that supports a wide range of carrier frequencies, deployment options, and use cases. NR can deliver very high data rates, e.g., in the order of multi-gigabit per second, due to the availability of larger bandwidth enabled by operation at higher frequency bands where carrier frequencies up to 52.6 GHz are supported in Release 15/16, and an extension to 71 GHz is being considered in Release 17. Subsequently, the terahertz bands (e.g., 100 GHz to 10 THz) are being considered as a potential enabler of ultra-high data rates in beyond 5G or sixth-generation (6G) networks~\cite{6G_Mzorzi,Thz_rappaport}. 

Though higher frequencies offer large chunks of the radio spectrum, high propagation loss at these frequencies necessitates the use of high antenna gain with narrow beam based directional transmissions. In such directional systems, since the spatial coverage for each \gls{tx} beam is limited, multiple beams are needed for transmitting  \gls{dl} common channels (e.g., system information, paging, etc.) to cover the entire cell area. In this paper, we focus on the paging procedure in these directional systems. 

To reduce UE's power consumption, a discontinuous reception (DRX) mechanism is used in NR (similar to LTE/LTE-A)~\cite{TS38304}. A UE's Radio Resource Control (RRC) connection is released and UE either enters an RRC IDLE or INACTIVE state when there is no scheduled data. The network then uses paging transmissions to inform the UE about any incoming calls/data, system information change, Earthquake and Tsunami Warning System (ETWS) notifications, or Commercial Mobile Alert Service (CMAS). A Paging message is transmitted over all the cells belonging to the list of Tracking Areas (in Core Network (CN)-initiated paging) or Radio-Network Areas (in RAN-initiated paging) for which the UE is registered. The UE is then configured to periodically wake up once in every DRX cycle (also known as paging cycle~\cite{TS38331}) to monitor if there is any paging message intended for the UE. 

In a directional NR system, a \gls{gnb} covers the entire cell by transmitting the same paging message over all the supported beams via beam-sweeping. The number of concurrent high gain beams that a \gls{gnb} can support may be limited by the cost and complexity of the utilized transceiver architecture. At high frequencies, the number of concurrent high gain beams supported in practice is much less than the total number of beams used to cover the cell area~\cite{MADP2018}. Therefore, paging transmissions take place over different time slots. As the carrier frequency increases, the number of beams required to cover the entire cell increases due to the higher beamforming gain required to overcome propagation loss limitations. This is a crucial challenge since the network's resource requirement for paging transmission would increase with the increase of the number of beams. For example, based on the analysis provided in~\cite{MADP2018} and~\cite{R2-168125}, the resource requirement may exceed system capacity ($>$100\%) to support high (e.g., 128 or higher) number of beams as the paging rate increases, making NR based paging procedure incapable of supporting highly directional systems. 

The paging resource overhead problem in directional systems has been considered in few earlier works~\cite{agiwal_sensors2018, mmw_paging_icc2018}. In~\cite{agiwal_sensors2018}, paging resource overhead is reduced using shorter UE IDs and the proposed mechanism is shown to achieve 15\% gain in \gls{gnb} power savings with 20\% shorter UE IDs. Authors in~\cite{mmw_paging_icc2018} proposed the use of different paging cycles for non-delay and delay-sensitive UEs, a short paging cycle using only the sub-set of beams for the delay-sensitive UEs, and a long paging cycle using all the beams for non-delay sensitive UEs. However, the proposed solution can only be supported in RAN-initiated paging where \gls{gnb} retains the UE context. 

In this paper, we present a minimal feedback enabled paging mechanism which can support both CN-initiated and RAN-initiated paging. In the proposed solution, instead of using all the beams for paging transmissions, only a sub-set of beams are used (activated) based on UE(s) presence. In every paging cycle, beam activation is based on the history of UEs presence under its coverage where a UE presence is indicated using a minimal set of resources. A similar mechanism is also proposed in~\cite{MADP2018} where UEs are configured to send presence indications in every paging cycle incurring significant energy burden at the UE side in order to reduce \gls{dl} paging overhead. Our approach curtails \gls{dl} paging overhead significantly while at the same time minimizes the energy cost at the UE side. The main contributions of this paper are summarized as follows:

\begin{itemize}
\item A proposal of a baseline paging solution that minimizes the UE power consumption associated with UE presence indications compared to the existing solution~\cite{MADP2018}.
\item An analytic model and derivation of an average number of active beams and UE presence indications over a defined duration for the baseline paging solution.
\item Further enhancements to the baseline paging solution proposal for more efficient UE power consumption associated with UE presence indications.
\item Extensive evaluation of the proposed solutions in terms of paging resource utilization and UE power consumption compared to legacy 3GPP NR solution and literature~\cite{MADP2018}.
\end{itemize}

The remainder of the paper is organized as follows. Section~\ref{sec:Nrpag} reviews the paging procedure for NR based directional systems. In Sec.~\ref{sec:prosol}, the proposed baseline minimal feedback paging solution, its analytic model, and further enhancements to reduce UE power consumption are presented. The analytic model verification and evaluation is then presented in Sec.~\ref{subsec:ana_eval}. The performance of the proposed solutions is evaluated in Sec.~\ref{sec:pereva}. Finally, conclusions are summarized in Sec.~\ref{sec:conc}.

\section{Paging in 3GPP NR based Directional Systems}\label{sec:Nrpag}
The network may configure a paging cycle with multiple \gls{po}s depending on the paging load. UEs are configured by the network with the paging cycle length, the number of paging frames in a paging cycle, and the number of \gls{po}s in a paging frame~\cite{TS38331}. Using such configuration along with the associated UE ID, a UE determines the paging frames and the \gls{po}s to be monitored~\cite{TS38304}. In this paper, a group of UEs monitoring the same \gls{po}s is referred to as a paging group. 

\begin{figure}[!t]
	\centering
	\includegraphics[width=0.5\textwidth]{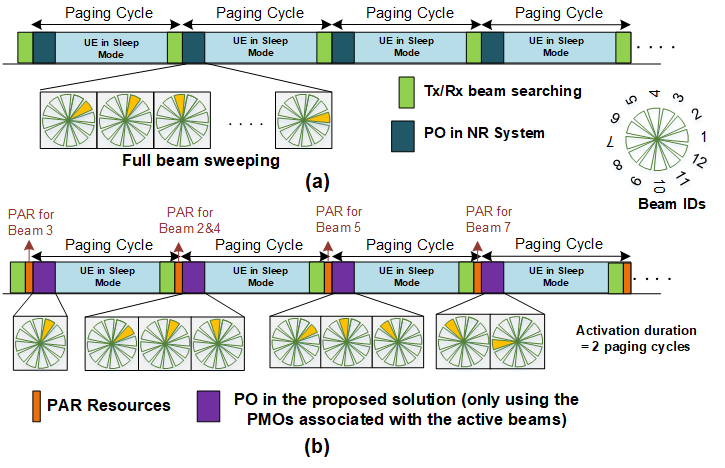}
	\captionsetup{justification=centering, font=small}
	\caption{Example of Paging transmissions in: (a) NR based solution using full beam sweeping, (b) Proposed solution, using only active beams, where after receiving a PAR, a beam gets activated for the activation duration.}
	\label{fig_proc_ex}
	\vspace{-0.6cm}
\end{figure}

In 3GPP NR, in directional operations, each \gls{po} is a set of \gls{pdcch} monitoring occasions (PMOs) and can consist of multiple time slots (e.g., subframe or OFDM symbol)~\cite{TS38304}. A paging transmission, which can be paging \gls{dci} consisting of either Short Message (for system information update, ETWS notification, and CMAS), scheduling information of the paging message over \gls{pdsch}, or paging message over \gls{pdsch}, is repeated in all \gls{dl} \gls{tx} beams by a \gls{gnb}. To enable that, in each \gls{po}, every supported \gls{dl} \gls{tx} beam is allocated with at least one dedicated PMO.  An example is shown in Fig.~\ref{fig_proc_ex}(a), where in each \gls{po}, paging \gls{dci} transmissions by a \gls{gnb} are performed over all the supported \gls{dl} \gls{tx} beams (or over the PMOs associated with all the supported \gls{dl} \gls{tx} beams). The UEs are configured with the parameters needed to identify the association between the PMOs and the \gls{gnb}'s \gls{tx} beams~\cite{TS38304}. To determine the PMO needs to be monitored in a \gls{po}, a UE is required to determine the \gls{dl} \gls{tx} beam over which it would receive the paging information. Thus, beam searching is incorporated in directional operations, as shown in Fig.~\ref{fig_proc_ex}(a). The UE can use periodic Synchronization Signal Blocks (SSBs) transmitted by the \gls{gnb} over all the supported \gls{dl} \gls{tx} beams, where the UE can wake up before its PO and measures the signal quality of SSBs from each of the \gls{gnb}'s \gls{tx} beam to determine the best \gls{dl} \gls{tx} beam. If the UE also supports beamforming, it can use different RX beams to measure the signals from the \gls{gnb} to identify the best RX beam as well. 

Note that in the remaining part of the paper, the term beam/beams refers to \gls{dl} \gls{tx} beam/beams.


\section{Proposed Minimal Feedback Enabled Paging}\label{sec:prosol}
We introduce a newly defined concept of a set of active beams to mitigate the increasing resource overhead problem associated with paging transmissions in directional systems. For any \gls{po}, a \gls{gnb} performs paging transmissions associated with the paging group of that PO only over a defined set of active beams. In order to identify the set of active beams, \gls{ul} feedback transmissions, called Paging Activation Requests (PARs) herein, initiated by UEs in IDLE/INACTIVE state are required. To enable PARs, beam and Paging group specific time-frequency resources (e.g., one-to-one mapping between a resource and a beam for each Paging group) are allocated. These resources can be allocated in every paging cycle with time resources preceding those of the \gls{po}s. 
 
PARs can be used by UEs to activate beams through simple illumination of associated resources and using energy detection at \gls{gnb}. Hence, a UE indicates its presence to the \gls{gnb} (where it is camped on) by sending a PAR over the allocated resource to the UE's best detected beam. On a successful energy detection over that resource, the \gls{gnb} determines that there is at least one UE present under the coverage of the associated beam, and therefore activates that beam for the \gls{po}s associated with the UE's paging group. Since \gls{gnb}s utilize non UE-specific and energy-based detection to activate beams based on received PARs, only a minimal resource allocation is required for PARs, for which an example is given in Sec.~\ref{sec:pereva}.  

As mentioned earlier, resources for PARs can be allocated in each paging cycle, however, sending PARs in every paging cycle can incur significant energy consumption at the UE. Therefore, we introduce the activation duration as another new concept. On the reception of a PAR, the \gls{gnb} activates the associated beam for the activation duration, defined in terms of a number, $N_a$, of Paging Cycles. An example procedure is shown in Fig.~\ref{fig_proc_ex}(b) for an activation duration of 2 paging cycles where paging \gls{dci} transmissions by a \gls{gnb}, in each \gls{po}, are performed only over the PMOs associated with the activated beams based on received PARs. At the UE side, after sending a PAR for a beam, the UE does not need to send a request for the same beam at least for the next activation duration.

\subsection{Analytic Modeling of Proposed Baseline Scheme}\label{sec:anamod}
We develop an analytic model to provide an insight into the performance of the aforementioned proposed baseline paging mechanism. In \gls{dl}, the proposed mechanism provides paging resource overhead reduction by activating only a sub-set of beams based on UE(s) presence compared to all the beams in the legacy NR system, therefore, we derive the average number of activated beams. Afterward, we derive the average number of PARs incurred by a UE which represents the additional UE's overhead associated with the proposed scheme. 

To determine the average number of dynamically and uniquely activated beams $\bar{n}$ over any period of $N_a$ paging cycles for a specific cell of circular radius $R_c$, we assume that UEs are distributed according to a homogeneous and stationary Poisson Point Process (PPP) with density~$\lambda$~(UEs/Cell), i.e., number of UEs follow a Poisson distribution and UEs' locations are uniformly distributed over the cell coverage. The average number $\bar{n}$ can then be obtained as 
\begin{equation}
\bar{n} = \sum_{i=1}^{N_a}\mathbb{E}\left[ n_i \right],
\end{equation}
where $n_i$ is a random variable representing the number of uniquely activated beams in the $i^{th}$ paging cycle which are not activated in any of the $j^{th},~j \in \{i+1, i+2, ..., N_a\}$ paging cycles. Due to the inter-dependency between the random variables $n_i,~i \in \{1, 2, ..., N_a\}$, we consider the following recursive formula to obtain the expected values of $n_i \forall i$ as
\begin{equation}\label{eq:nestexpect}
\begin{aligned}
    &\mathbb{E}\left[ n_i \right] = \\
    & \mathbb{E}_{u_i, n_j, u_j}\left[ \mathbb{E}_{n_i | u_i, n_j, u_j}\left[ n_i \right]  \right],~j \in \{i+1, i+2, ..., N_a\},
\end{aligned}
\end{equation}
where $u_i$ is a random variable representing the number of UEs under the coverage of $n_{u,i} = B_{TX} - \sum_{j=i+1}^{N_a} n_j$ of beams that are not activated at any of the paging cycles $j \in \{i+1, ..., N_a\}$ for a total number of $B_{TX}$ beams at the \gls{gnb} and $n_{u,N_a} = B_{TX}$. A conditional and a joint probability mass functions (PMFs) are required for the evaluation of \eqref{eq:nestexpect} which can be expressed, respectively, as
\begin{eqnarray}
\mathbb{P}\{n_i|u_i, n_j, u_j\} &\!\!=\!\!& \frac{\left(\!\!\begin{array}{c}
    n_{u,i}  \\
     n_i 
\end{array}\!\!\right)}{n_{u,i}^{u_i}} b_{n_i},\;\; 1 \leq n_i \leq \min \{u_i, n_{u,i}\}, \nonumber\\
\text{where,\;\;} b_{n_i} &\!\!=\!\!& n_i^{u_i} - \sum_{k=1}^{n_i - 1} \left(\!\! \begin{array}{c}
     n_i  \\
     k 
\end{array} \!\!\right) b_k, \;\; b_1 = 1,  \label{eq:CondProb}\\
\mathbb{P}\{u_i, n_j, u_j\} &\!\!=\!\!& \mathbb{P}\{u_i|n_j, u_j\} \mathbb{P}\{n_j, u_j\} \nonumber\\
&\!\!=\!\!& \frac{\lambda_i^{u_i} e^{-\lambda_i}}{u_i !}\mathbb{P}\{n_j, u_j\}, \label{eq:JointProb}
\end{eqnarray}
where the expression in \eqref{eq:JointProb} follows from the PPP distribution assumption, $\lambda_i = \lambda n_{u,i}/B_{TX}$, and $\mathbb{P}\{u_{N_a}, n_{N_a+1}, u_{N_a+1}\} = \mathbb{P}\{u_{N_a}\}={\lambda_i^{u_{N_a}} e^{-\lambda}}/{u_{N_a} !}$. The joint distribution $\mathbb{P}\{n_j, u_j\} = \mathbb{P}\{ n_{i+1}| u_{i+1}, n_l, u_l \} \mathbb{P}\{u_{i+1}, n_l, u_l\},~l \in \{i+2, i+3, ..., N_a\}$ can then be recursively evaluated according to the conditional and joint PMFs presented in \eqref{eq:CondProb} and \eqref{eq:JointProb}.

Then, for a specific paging group, we can express the total number of network resources utilized for paging in NR based solution as $R_D \times B_{TX}$, where $R_D$ represents the total number of resources utilized per beam over the $N_a$ paging cycles. On the other hand, the total network resources utilized considering the proposed solution can be expressed as $\bar{n} \times R_D$. To account for UL resources required to send PARs, we use $R_U$ to represent the number of resources required for a single PAR transmission and express the total number of resources over the $N_a$ paging cycles for the specific paging group as $N_a \times R_U \times B_{TX}$. We can then compare the total number of required resources by the proposed scheme to the NR based solution using the gain factor $\Gamma$ defined as
\begin{equation}
\Gamma = 1 - \frac{(R_D \times \bar{n} + R_U \times B_{TX} \times N_a)}{(R_D \times B_{TX})}.
\end{equation}

As mentioned earlier, due to energy-based detection for PARs at the gNB and therefore the dominance of the DL resources $R_D$ over UL resources $R_U \times N_a$, the gain factor $\Gamma$ will majorly vary with $\bar{n}/B_{Tx}$. 

Additionally, we can evaluate the cost of UE's power consumption incurred by the PAR transmissions associated with the proposed solution. The average amount of energy expended by a UE on PAR transmissions over a period of $N_a$ paging cycles will be linearly proportional to the average number of PAR transmissions ($\bar{k}$). Since in the proposed solution a beam is activated for $N_a$ paging cycles following the detection of a corresponding PAR, $\bar{k}$ for a particular UE will depend on the number of occurrences the UE switches its beam, which can be determined as 
\begin{eqnarray}\label{eq:uln}
\bar{k} = 1 \!+\!\! \sum_{j=1}^{N_a-1} j {N_a-1 \choose j} \left(1-\frac{1}{B_{Tx}}\right)^{j} \! \left(\frac{1}{B_{Tx}}\right)_.^{N_a-1-j} \!\!\!\!\!\!\!\!\!\!
\end{eqnarray}

Note that, in the above calculation of $\bar{k}$, for simplicity of calculation, we assume that a UE can only keep track of activation duration of maximum one beam at a time, however, for performance evaluation using simulations in Sec.~\ref{sec:pereva}, we will consider the scenario where a UE can keep track of the activation duration of multiple beams.

\subsection{Enhancements for Minimizing UE's Energy Expenditure}\label{sec:enhsol}
In high-frequency bands, a highly directional system is required leading to higher levels of power consumption. For example, using the demonstrated 80 to 100GHz phased-array transceiver in \cite{Shahramian_phased_array}, a transceiver that employs 64 antennas or more may result in an estimated transmitter power of $1.1$W. 
Therefore, we introduce enhancements to the proposed baseline paging scheme that further minimizes the number of PAR transmissions required by a particular UE. Two of those enhancements include \textit{DL Indication of Active Beams} and \textit{Monitoring Duration} at UE. 

\textbf{\gls{dl} Indication of Active Beams}: when a \gls{gnb} activates (i.e., activating an inactive beam) or re-activates (i.e., re-initiating the activation duration counter of an active beam) one or more beams based on the received PARs, the \gls{gnb} sends a \gls{dl} message, over all of the currently active beams, containing information about the (re-)activated beams. This enables UEs to minimize PAR transmissions by tracking the active beams associated with their paging group. Therefore, when a UE transitions to a beam that has been recently activated as indicated by a received \gls{dl} indication message of active beams, it does not need to send a PAR at least for the activation duration after receiving the corresponding \gls{dl} indication message. Existing paging \gls{dci} message (with or without Short Message or/and scheduling information for paging message) can be used to send such \gls{dl} active beam indications, e.g., a list of beam indices (e.g., associated SSB-indices as used in 3GPP NR) of the (re-)activated beams or a bit-map with a length equal to the total number of supported \gls{gnb} \gls{tx} beams where the bits corresponding to (re-)activated beams may be set to '1', otherwise set to '0', can be used.

\textbf{Monitoring Duration}: when a UE determines the need to activate a beam (e.g. transitions to a beam for which it has not received an indication in a DL active beam indication message), it may first monitor the beam for a configured monitoring duration which may be defined as an integer number, $N_m$, of paging cycles. If the UE detects any paging \gls{dci} over that beam during the monitoring duration, then it determines that the beam has already been activated without the need to transmit a PAR. Otherwise, the UE transmits a PAR to request the activation of that beam. This empowers a UE with an ability to take advantage of other UEs' PAR transmissions that have already activated the desired beam. 

UEs with different mobility states may be configured with different values of monitoring duration, $N_m$. The mobility states for an IDLE/INACTIVE state UE can be defined at the cell level, e.g. as a function of cell re-selection rate as defined in \cite{TS38304}, or at a beam level, e.g. as a function of beam re-selection rate, for highly directional systems. A UE with high mobility (i.e., high beam-changing/re-selecting rate) may be configured with a lower value of monitoring duration as compared to another UE with low mobility. This is to avoid the situation, for example, when a UE moves fast enough such that the UE's best detected beam is changed within the monitoring duration. In this case, the UE may not get the opportunity to activate the desired beam if needed, which can subsequently incur significant paging latency for that UE. Paging latency for a particular UE can be defined as the total time from paging request arrival to the successful transmission of the paging message to the UE. 

\begin{figure}[!t]
	\centering
	\includegraphics[width=0.45\textwidth]{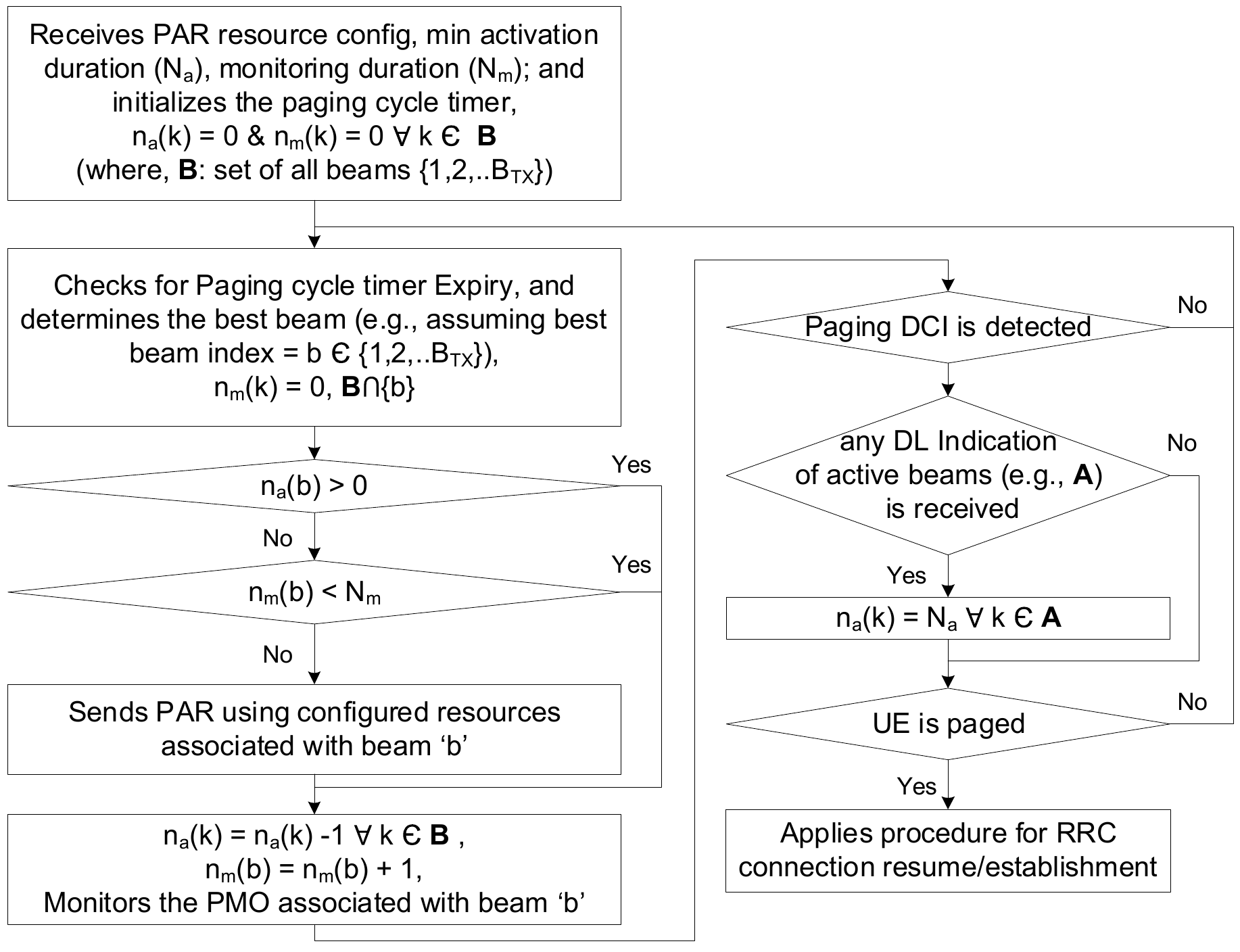}
	\captionsetup{justification=centering, font=small}
	\caption{UE Procedure to enable the proposed paging solution.}
	\label{fig_uep}
	\vspace{-0.4cm}
\end{figure}

Different paging solutions may be enabled by incorporating either or both \gls{dl} indication of active beams and monitoring duration along with the activation duration concept described earlier in this section. In Fig.~\ref{fig_uep}, a UE procedure is shown, where, both \gls{dl} indication of active beams and monitoring duration are enabled. Next, we summarize the State of Art (SoA) and proposed paging solutions in Sec.~\ref{subsec:PagingSolutionsSummary} and compare the performance of the presented solutions in Sec.~\ref{sec:pereva}.   

\subsection{Summary of SoA and Proposed Paging Solutions}\label{subsec:PagingSolutionsSummary}
In this section, we present the set of paging solutions that will be subject to the evaluation and comparison in Sec.~\ref{sec:pereva}. These range from the legacy standardized \cite{TS38304} to SoA \cite{MADP2018} to proposed solutions in \ref{sec:enhsol} which are described as
\begin{itemize}
\item \textbf{Legacy}: current standardized solution in 3GPP NR specification~\cite{TS38304} described in Sec.~\ref{sec:Nrpag}.
\item \textbf{MADP}: SoA solution in\cite{MADP2018} that minimizes DL resources overhead, where a UE sends a PAR associated with its best beam in every paging cycle, and the \gls{gnb} activates all the beams for which at least one PAR is received.   
\item \textbf{MFEP-AD}: proposed baseline scheme modeled in Sec.~\ref{sec:anamod} with a configured beam activation duration and without the enhancements described in Sec.~\ref{sec:enhsol} (i.e., \gls{dl} indication of active beams and monitoring duration).
\item \textbf{MFEP-DLI}: MFEP-AD solution augmented with the proposed enhancement of \gls{dl} indication of active beams described in Sec.~\ref{sec:enhsol}.
\item \textbf{MFEP-MD}: MFEP-DLI solution augmented with the proposed enhancement of paging monitoring duration configuration described in Sec.~\ref{sec:enhsol} and shown in Fig.~\ref{fig_uep}.    
\end{itemize}


\section{Analytic Model Verification and Evaluation}\label{subsec:ana_eval}
In this section, we verify the performance of analytic models developed in Sec.~\ref{sec:anamod} against simulation. We also use the analytic model to evaluate the performance of the proposed baseline scheme in terms of the average number of activated beams for paging and PAR transmissions. For verification, we use Monte Carlo simulation to generate 10,000 random realizations of UEs distribution within the coverage of a cell of radius $R_c$ = 100m where each realization corresponds to a single paging cycle with UEs assumed to be stationary.

\begin{figure}[!t]
	\centering
	\includegraphics[width=0.49\textwidth]{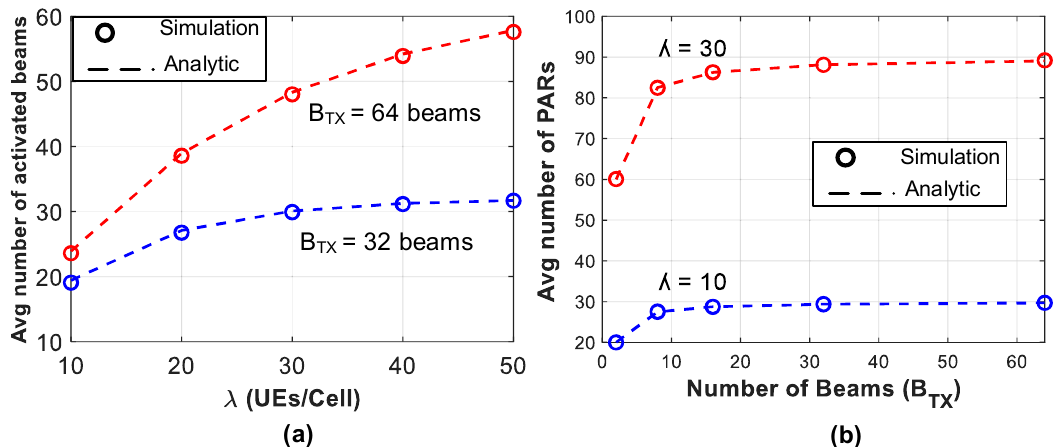}
	\captionsetup{justification=centering, font=small}
	\caption{Results of the analytic model in Sec.~\ref{sec:anamod}: (a) Average number of activated beams (b) Average number of PARs, over $N_a$ = 3 paging cycles.}
	\label{fig_anaRes}
	\vspace{-0.4cm}
\end{figure}

Fig.~\ref{fig_anaRes}(a) compares analytic model results to simulation and shows the average number of activated beams ($\bar{n}$) over $N_a$ = 3 paging cycles, with different UE density ($\lambda$) and total number of supported beams ($B_{Tx}$) per cell. As shown in the figure, the results with the analytic expressions in Sec.~\ref{sec:anamod} closely match the simulation results. Further, as expected, the average number of beams activated by at least one UE increases with the number of UEs. We can also note that for a given UE density, as the total number of supported beams per cell increases (or equivalently as the beam's coverage area reduces), the average number of beams activated by at least one UE will also increase. 
Most importantly, it can be concluded based on the result shown in Fig.~\ref{fig_anaRes}(a) that as long as the number of UEs within a cell coverage is being less than the total number of supported beams in that cell, the proposed baseline solution in Sec.~\ref{sec:anamod} will require a significantly smaller number of beams to be activated for paging in contrast to the total number of supported beams in the legacy NR system. This can be translated into a significant resource utilization reduction compared to the legacy NR system. 

On the other hand, Fig.~\ref{fig_anaRes}(b) shows the average total number of PARs generated by all UEs within a cell coverage over $N_a$ = 3 paging cycles according to the analytical model presented in Sec.~\ref{sec:anamod}. Due to the lower probability of a UE staying under the coverage of the same beam over independent location distribution realizations that are associated with the increase in the number of beams, we note an increase in the average total number of PAR transmissions. Additionally, as one would expect, the average total number of PAR transmissions increases with the increase in UE density. Please note that these results are based on the analytical model where we assume that a UE can keep track of the activation duration of only one beam, once we relax this constraint and also enable the enhancements described in the Sec.~\ref{sec:enhsol}, we will show next in the simulation-based results in Sec.~\ref{sec:pereva} that the average number of PAR transmissions can be significantly lower.      

\section{Performance Evaluation}\label{sec:pereva}\label{subsec:detSim_res}
In this section, we provide an extensive evaluation of performance for the paging solutions summarized in Sec.~\ref{subsec:PagingSolutionsSummary} using Monte Carlo simulations. 

\subsection{Simulation Assumption}\label{subsubsec:sim_assump}
In our simulations, we consider a system shown in Fig.~\ref{fig_SimResvarpr}(a) comprising a tracking area consisting of 16 gNBs in a dense urban scenario with an inter-site distance of 200m~\cite{R2-165202}. We further consider different UE densities defined as the number of UEs within a paging group (i.e., supported per PO). A UE density can then be translated into the number of UEs paged per second based on the selection of system parameters such as the paging cycle, number of paging frames per paging cycle, and number of POs per paging frame. 
Additionally, UEs are initially randomly dropped within the simulated tracking area where we assume that 40\% of UEs are stationary, another 40\% have low mobility (i.e., speed of 3km/hr), and the remaining 20\% have high mobility (i.e., speed of 30km/hr); and random walk is considered as the mobile UEs' mobility model. 

The 3GPP FTP traffic model is considered to generate paging requests per UE according to a Poisson distribution with an average arrival rate ($\lambda_p$) of 1 packet per 60 seconds~\cite{hybrid_paging2017}. A system bandwidth of 400 MHz and sub-carrier spacing of 120 KHz are considered, as recommended by 3GPP for millimeter-wave frequency bands\cite{TS38101_2}, which correspond to a total of 264 available RBs for any DL/UL transmission. A paging \gls{dci} is transmitted using Control Resource Set 0, which uses one OFDM symbol and 48 resource blocks (RBs)~\cite{TS38213}. The \gls{pdsch} paging messages containing the paged UEs' IDs are assumed to have the following configuration: a 48 bit UE ID~\cite{TS38304}, QPSK modulation, and 0.37 code rate~\cite{TS38214}. \gls{dl} indication of active beams associated with MFEP-DLI and MFEP-MD solutions is assumed to be sent via a DCI with bit-map information as mentioned in Sec.~\ref{sec:enhsol}. The considered DCI occupies one OFDM symbol and a number of frequency domain resources (e.g., RBs) determined based on the total number of supported beams. According to 3GPP specifications~\cite{TS38331}, the considered DCI requires a total number of \{6, 6, 6, 12, 24\} RBs, corresponding to a total number of \{108, 108, 108, 216, 432\} coded bits, to support a total number of \{16, 32, 64, 128, 256\} beams, respectively.


For the MFEP-MD scheme, different values of monitoring duration are configured for different UE mobility states, as mentioned in Sec.~\ref{sec:enhsol}. In our simulations, we consider two different configurations: MFEP-MD (4/2/0) and MFEP-MD (6/3/0) where the values (x/y/z) represent the monitoring duration defined in the number of paging cycles corresponding to UEs with \{no, low, high\} mobility states, respectively.

For PAR transmissions, each transmission utilizes minimal time and frequency resources due to the energy-based detection considered at the \gls{gnb}~\cite{MADP2018}. In our simulations, we consider a single resource element in the frequency domain and two OFDM symbols in the time domain which is the minimum number of symbols allocated for a random access preamble transmission in NR for millimeter-wave frequency bands such that interference to/from other control/data transmissions can be avoided. The remaining parameters considered in the simulations are provided in Table~\ref{tab:simPa}.

\begin{table}
\centering
\footnotesize
\begin{tabular}{|m{5.6cm}|m{2.1cm}|} 
 \hline
   \textbf{Parameter} & \textbf{Value} \\
 \hline
 Paging Cycle Duration & 320ms~\cite{TS38331} \\
 \hline
 Total number of Paging Cycles & 100,000 \\
 \hline
 Activation Duration ($N_a$) & 5 paging cycles \\
 \hline
 Max number of UEs can be paged simultaneously & 32~\cite{TS38331} \\
 \hline
\end{tabular}
\caption{Simulation Parameters}
\label{tab:simPa}
\vspace{-0.4cm}
\end{table}

\subsection{Simulation Results}\label{subsubsec:detailed_sim_res}

\begin{figure}[!t]
	\centering
	\includegraphics[width=0.5\textwidth]{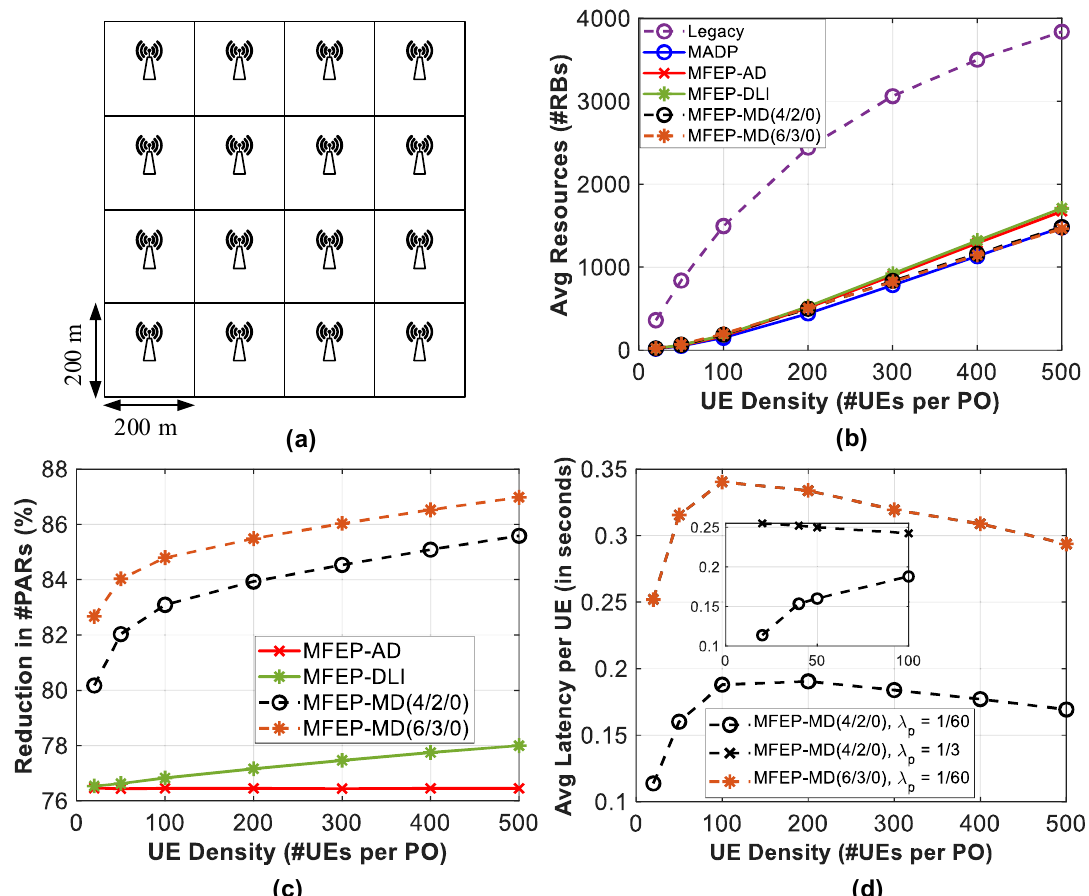}
	\captionsetup{font=small}
	\caption{(a) Simulations system setup with $B_{TX} = 64$, (b) Average number of resources (in \#RBs) per paging cycle per cell, (c) Reduction (\%) in number of PARs by proposed solutions compared to MADP~\cite{MADP2018}, (d) Average Latency incurred by the MFEP-MD based solutions.}
	\label{fig_SimResvarpr}
	\vspace{-0.4cm}
\end{figure}

We first present the results for a total fixed number of 64 supported beams per cell while varying UE density. Fig.~\ref{fig_SimResvarpr}(b) shows the average number of network resources utilized (for both \gls{dl} paging-related and \gls{ul} PAR transmissions) per paging cycle per cell for different solutions. We note a significant reduction in paging-related resource utilization for the MADP and the proposed solutions compared to the Legacy solution. This gain is a result of the utilization of a much lower number of beams for paging transmissions compared to the all 64 beams in the Legacy system. We also note that all the proposed solutions utilize approximately the same amount of resources for paging as the MADP scheme for the simulated range of UE densities. For example, paging resource utilization by the MADP and the proposed solutions correspond to a range of 80-81\% reduction in resources compared to the Legacy solution at a UE density of 200 UEs/PO. 

On the other hand, we show in Fig.~\ref{fig_SimResvarpr}(c) that our proposed solutions result in a much lower number of PAR \gls{ul} transmissions compared to the MADP solution which makes the proposed solutions more favorable in terms of UE energy consumption. We also note that the number of PARs in the MFEP-DLI and the MFEP-MD solutions decrease further as the UE density increases which are a result of the utilization of \gls{dl} indication of active beams and monitoring duration that allow UEs to benefit from the already activated beams by other UEs.
We further note that the configured monitoring duration in the MFEP-MD solution, which forces a UE to first monitor a beam before deciding to transmit a PAR, enables to provide more reduction in \gls{dl} resource utilization and UE energy consumption compared to the MFEP-DLI and the MFEP-AD solutions. The reduction in the number of PAR transmissions for the proposed solutions comes at a cost of either a slight increase in \gls{dl} resource overhead for the MFEP-DLI solution compared to the MADP and the MFEP-AD solutions, i.e. due to the transmission of \gls{dl} indications of active beams, or as discussed below, an increase in paging latency (discussed next) for the MFEP-MD solution compared to other solutions, i.e. due to the configured monitoring duration. 


We now show in Fig.~\ref{fig_SimResvarpr}(d) the latency incurred by the MFEP-MD solution. We first note that increasing the monitoring duration, i.e., MFEP-MD(6/3/0) vs. MFEP-MD(4/2/0), increases the paging latency experienced by UEs for simulated UE densities, but on the other hand results in more reduction of \gls{dl} resource utilization and UE energy consumption as shown in Figs.~\ref{fig_SimResvarpr}(b) and~\ref{fig_SimResvarpr}(c).  
We then note that, for the considered arrival rate of paging requests ($\lambda_p$ = 1 packet per 60 seconds, i.e., 1/60), paging latency first increases with the UE density until a point when UEs can start taking advantage of the activated beams by other UEs. However, the upward trend at low UEs density vanishes as we increase the $\lambda_p$ to 1/3 as shown in Fig.~\ref{fig_SimResvarpr}(d). This suggests that the MFEP-MD solution may be preferred for a specific combination of UE densities and arrival rate of paging requests. Further, for the simulated range of UE densities, all other solutions do not experience paging latency. However, this might not be the case for a high enough UE density or/and arrival rate of paging requests when the network might not be able to page all the UEs at the paging cycle within which the associated paging requests arrive.

\begin{figure}[!t]
	\includegraphics[width=0.5\textwidth]{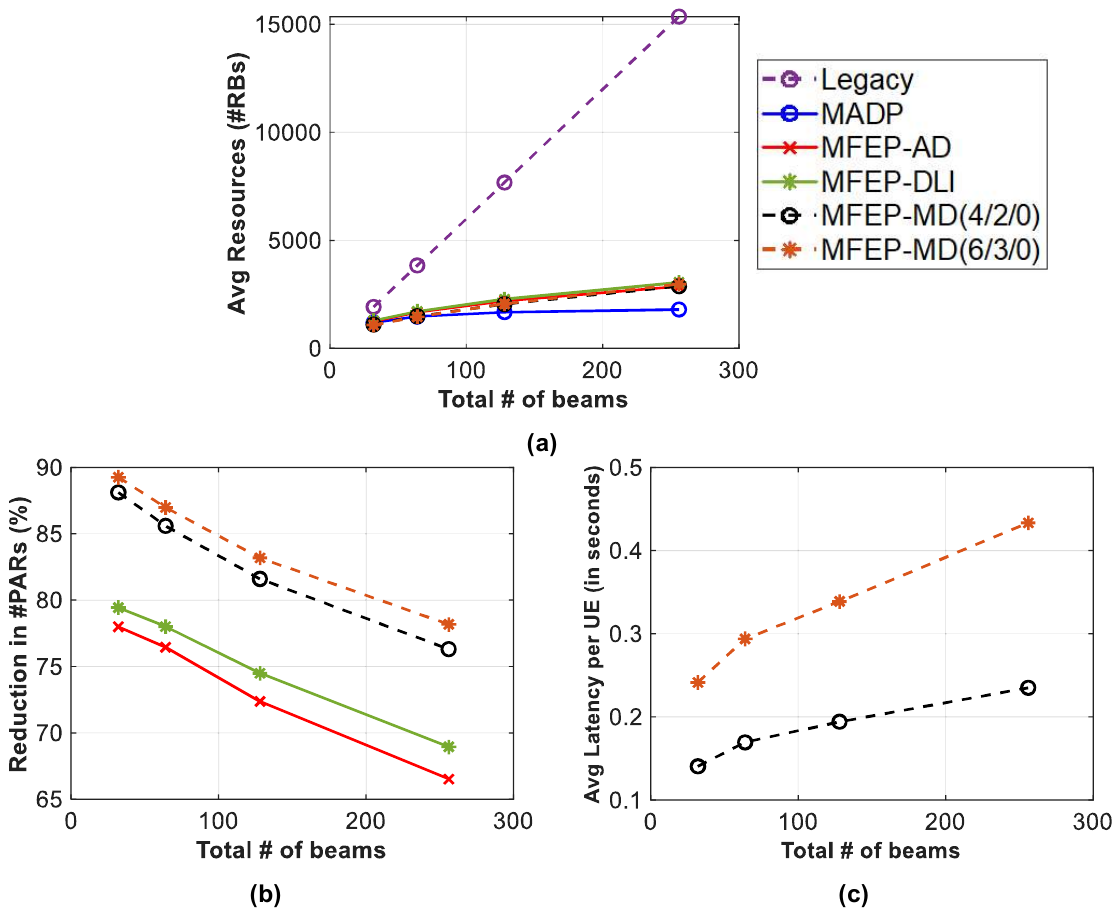}
	\captionsetup{font=small}
\caption{Results with UE density of 500 UEs/PO and varying number of total beams: (a) Average number of resources (in \#RBs) per paging cycle per cell, (b) Reduction (\%) in number of PARs by proposed solutions compared to MADP~\cite{MADP2018}, (c) Average Latency incurred by the MFEP-MD based solutions.}
	\label{fig_SimResvarbeams}
	\vspace{-0.4cm}
\end{figure}

Next, we present in Fig.~\ref{fig_SimResvarbeams} the results for a fixed UE density of 500 UEs/PO while varying the total number of supported beams. 
Fig.~\ref{fig_SimResvarbeams}(a) shows that resource utilization for paging in the Legacy solution increases significantly and linearly with the increase in the number of supported beams, whereas for the MADP and the proposed solutions, the increase is slight and eventually the required resources saturate. 
Therefore, the resource utilization reductions using the MADP and the proposed solutions compared to the the Legacy solution becomes more prominent as the number of beams increases. We note that in general the amount of resources in the MADP and the proposed solutions are dependent on both the UE density and the number of beams required to cover the UEs within a specific cell. Subsequently and despite the fixed UE density, we observe a slight increase in the resource utilization by the MADP and the proposed solutions with the increase in the number of supported beams due to the reduction of the beam coverage area.
An additional reason for the increase in resource utilization for the MFEP-DL and the MFEP-MD solutions is the increase in the total number of bits required to send a DL indication of active beams. 

Further, we note that there is a slight increase in the number of utilized resources by the proposed solutions compared to the MADP solution as the number of supported beams increases. This is again due to the reduction in beams' coverage areas, which subsequently lead to a higher rate of beam switching for the mobile UEs. The higher rate of beam switching results in scenarios when a beam remains unnecessarily activated due to the configured activation duration while there are not any UEs to be served under its coverage. Therefore, the beam activation duration should take into account both the mobility states of served UEs as well as the supported beam's coverage area, i.e. number of supported beams per cell. 

Another result of the increased beam switching rate for the mobile UEs with the increase in the total number of supported beams is shown in Fig.~\ref{fig_SimResvarbeams}(b), where we note a lower reduction in the number of PARs for the proposed solutions compared to the MADP solution. However, it is still clear that the number of PARs transmitted by the proposed solutions is significantly lower than those required by the MADP solution, even with the high number (e.g., 256) of beams.  

For the performance of the MFEP-MD solution, a similar trend is observed compared to the other paging solutions as described earlier, where the higher reduction in the number of PARs is achieved at the cost of additional paging latency as shown in Fig.~\ref{fig_SimResvarbeams}(c). We also note that the paging latency for the MFEP-MD solution increases as the number of supported beams increases and that the latency is usually lower for a lower monitoring duration. This is also a result of the increase in the rate of beam switching corresponding to the UEs' mobility and the beam coverage area, and the configured monitoring duration as described earlier. Therefore, monitoring duration configuration should take into account the expected beam switching rate.  

\section{Conclusions}
\label{sec:conc}
In this paper, we considered the resource overhead problem associated with paging in highly directional systems based on the legacy 3GPP NR solution. We proposed few variants to a minimal feedback based paging solution, which can significantly reduce the paging resource overhead for reasonable UE densities by activating the sub-set of beams that provide coverage to IDLE/INACTIVE UEs based on their presence indicated by PARs. The MADP\cite{MADP2018} and the proposed solutions manifest more than 80\% reduction in paging resources compared to legacy 3GPP solution in a system supporting 64 beams per \gls{gnb} and at a UE density of 200 UEs/PO. On the other hand, the proposed solutions incur significantly lower energy consumption at the UE compared to the MADP solution.
Finally, we note that the network's performance can be optimized to enable efficient paging resource utilization at reasonable energy consumption and paging latency experienced by a served UE through a careful choice of the activation and monitoring duration configuration. The choice of such configuration, which will be considered in our future work, should be dependent on the total number of supported beams per cell, the UE density, the paging requests arrival rate, and as well as the UEs' mobility states.

\bibliography{references}
\bibliographystyle{ieeetr}

\end{document}